# Observation of re-entrant correlated insulators and interaction driven Fermi surface reconstructions at one magnetic flux quantum per moiré unit cell in magic-angle twisted bilayer graphene

Ipsita Das[1], Cheng Shen[1], Alexandre Jaoui[1], Jonah Herzog-Arbeitman[2], Aaron Chew[2], Chang-Woo Cho[3], Kenji Watanabe[4], Takashi Taniguchi [5], Benjamin A. Piot[3], B. Andrei Bernevig[2] and Dmitri K. Efetov[1]*

1.  ICFO - Institut de Ciencies Fotoniques, The Barcelona Institute of Science and Technology, Castelldefels, Barcelona, 08860, Spain
2.  Department of Physics, Princeton University, Princeton, New Jersey 08544, USA
3.  Laboratoire National des Champs Magnétiques Intenses, Univ. Grenoble Alpes, UPS-INSA-EMFL-CNRS-LNCMI, 25 avenue des Martyrs, 38042 Grenoble, France
4.  Research Center for Functional Materials, National Institute for Materials Science, 1-1 Namiki, Tsukuba 305-0044, Japan
5.  International Center for Materials Nanoarchitectonics, National Institute for Materials Science, 1-1 Namiki, Tsukuba 305-0044, Japan

*E-mail : dmitri.efetov@icfo.eu

**The discovery of flat bands with non-trivial band topology in magic angle twisted bilayer graphene (MATBG) has provided a unique platform to study strongly correlated phenomena including superconductivity, correlated insulators, Chern insulators and magnetism. A fundamental feature of the MATBG, so far unexplored, is its high magnetic field Hofstadter spectrum. Here we report on a detailed magneto-transport study of a MATBG device in external magnetic fields of up to $B = 31$ T, corresponding to one magnetic flux quantum per moiré unit cell $\Phi_0$. At $\Phi_0$, we observe a re-entrant correlated insulator at a flat band filling factor of $\nu = +2$, and interaction-driven Fermi surface reconstructions at other fillings, which are identified by new sets of Landau levels originating from these. These experimental observations are supplemented by theoretical work that predicts a new set of 8 well-isolated flat bands at $\Phi_0$, of comparable band width but with different topology than in zero field. Overall, our magneto-transport data reveals a qualitatively new Hofstadter spectrum in MATBG, which arises due to the strong electronic correlations in the re-entrant flat bands.**

Two superimposed graphene monolayers, misaligned by a small twist angle $\theta = 1.1°$, produce 8 electronic flat bands near the charge neutrality point [1]. The extreme flatness of these bands provides an exciting platform to study exotic quantum phenomena by pushing the particles into the strong coupling regime. A succession of new states, such as correlated insulators [2–5], superconductors [3, 4, 6–8], magnets [3, 9–11] and others, arise upon electrostatic doping of these flat bands. The moiré superpotential in magic angle twisted bilayer graphene (MATBG) creates a new set of renormalized bands in a mini-Brillouin zone which possess $C_{2z}$, $C_{3z}$, $C_{2x}$ rotational symmetries and time reversal symmetry $T$ at zero magnetic field. Non-zero Dirac helicity in MATBG, protected by the $C_{2z}T$ symmetry and the decoupled valleys, gives rise to the non-trivial band topology [12, 13, 14]. As the role of the symmetries in the observed quantum phases in MATBG is still a topic of active investigation, discovering novel flat band systems with different inherent symmetries, but similar phenomenology, is a major goal in the field.

In parallel, our exact theoretical study of the Bistritzer-MacDonald (BM) Hofstadter spectrum has uncovered a set of low energy flat bands (which are not Landau levels) persisting at one quantum of magnetic flux per moiré unit cell $\Phi_0$, where translation symmetry is restored, pushing the system into the strongly correlated regime [15]. While strong magnetic fields tend to disrupt the bands by breaking them into fractal Hofstadter sub-bands, here due to the Aharonov-Bohm effect, full density Bloch-like bands re-emerge at $\Phi_0$. Near the magic angle, the moiré unit cell area of MATBG is enlarged by a factor of $1/\theta^2$, which is about 3000 times larger than the graphene unit cell. Because of this substantial increase in size, the magnetic field required to reach one full flux quantum is far smaller, close to $B \sim 30$ T, and is within experimental reach [16, 17]. Although previous studies of the electronic properties in graphene/hBN superlattices in the presence of high magnetic fields have provided great insight into the Hofstadter spectrum [18-24], flat electronic bands with strong correlations have never been explored in this light before. Together, these characteristics make MATBG an unprecedented platform to study the Hofstadter spectrum enriched by (possible) band topology and strong interactions.

In this manuscript we report on the detailed magneto-transport behavior of a MATBG device with a twist angle $\theta = 1.12°\pm0.02°$ in the presence of an external perpendicular magnetic field as high as $B = 31$ T. This allows us to resolve the continuous evolution of the Hofstadter spectrum from zero field to one flux quantum per moiré unit cell $\Phi_0$, which corresponds to a $B$-field of $B_0 = 30.5$ T. Similar to the case of $B = 0$ T, at $B_0$ we observe re-entrant correlated insulators at certain integer fillings and interaction-driven Fermi surface reconstructions, evidenced by new sets of Landau levels originating from these. We also study the higher energy passive bands, which are theoretically predicted to be highly flattened at $\Phi_0$ compared to their zero-field dispersion, where we observe a rich interaction-reconstructed Hofstadter spectrum.

Fig. 1a shows the schematic of the device which consists of a van der Waals heterostructure of graphite/hBN/MATBG/hBN, where the insulating hexagonal boron nitride (hBN) layers were specifically non-aligned with the MATBG. We performed four terminal longitudinal resistance $R_{xx}$ and transverse resistance $R_{xy}$ measurements where the carrier density $n$ of the band is continuously tuned by a gate voltage $V_g$, which is applied to the graphite gate. The total carrier density $n$ is normalized by $n_S = 4n_0$, where $n_S$ is the density of the fully filled spin and valley degenerate moiré bands and $n_0$ is the density per flavor. The filling factor of the carriers per moiré unit cell is defined as $\nu = n/n_0$.

Fig. 1b illustrates the band structure of MATBG with $\theta = 1.12°$ at zero flux and at $\Phi_0$ flux where the bands have the same density. The zero-flux band structure has been intensely studied before [1, 12, 25] and consists of two connected nearly flat topological bands and higher energy dispersive bands. An exact study of the BM Hamiltonian in magnetic flux predicts the emergence of a set of low-energy flat bands at a flux of $\Phi_0 = 2\pi$, as studied in more detail in [15], which finds two very flat bands but with different symmetry and topology. The appearance of full-density flat bands at $\Phi_0$ flux suggests that the system will be dominated by strong interactions.

Fig. 1d shows $R_{xx}$ vs. $\nu$ measurements taken at a temperature of $T = 40$ mK in $B = 0$ T and $B = 30$ T, which corresponds to zero and one flux quantum $\Phi_0$ per moiré unit cell, respectively. The $B = 0$ T trace is dominated by a well-known sequence of resistance peaks, as has been reported in detail before in [2, 3]. At the charge neutral point (CNP) at $\nu = 0$, there appears to be a gapless phase since the resistance has a value on the order of $R_{xx} \sim 10$ k$\Omega$, as well as a set of correlated insulators (CI) at filling factors of $\nu = \pm2, +3$, and band insulators (BI) at $\nu = \pm4$. In addition, on the hole doping side of $\nu = -2$, we also observe a superconducting region (SC), where the resistance drops to zero (also see Extended data E). In the $B = 30$ T trace, we observe an overall similar picture, which strikingly shows enhanced resistance peaks at $\nu = 2$, indicative of re-entrant CIs. There are however also differences, as we do not observe clear signatures of CIs at other $\nu$ in this sample. Also, the CNP is now gapped with the resistance exceeding $R_{xx} > 10$M$\Omega$, and the gaps of the BIs at $\nu = \pm4$ are enhanced. We also do not find signatures of superconductivity, although we comment on this later.

To highlight the re-entrant behavior of the $\nu = 2$ CIs in $\Phi_0$ flux, we show the evolution of the $R_{xx}$ vs. $\nu$ in $B$-field in the color plot of Fig. 1c. The $\nu$-$B$ phase space appears to be nearly symmetric about the point corresponding to $\nu = 2$ and $B = 15$ T $\approx 0.5$ $\Phi_0$, under a symmetry sending $\nu$ to 4-$\nu$, and $B$ to $\Phi_0$-$B$. This symmetry is approximate and only holds well close to half-flux and filling $\nu = 2$. The dark red region at $\nu = 2$ shows the CI state at $B = 0$ T, originating from which we observe a set of Landau levels gaps (LLs) with LL filling factors $\nu_L = +2, +4$ etc., where the $\nu_L = +2$ was previously interpreted as a correlated Chern insulator with a Chern number $C = 2$ [26]. The corresponding $R_{xx}$ and $R_{xy}$ vs. $\nu$

line traces for $B = 5$ T are plotted in Fig. 1f, and show clear quantum Hall signatures of $\nu_L = +2$ with $R_{xx} \sim 0\ \Omega$ and $R_{xy} \sim 12.5$ kΩ. The resulting electron degeneracy of 2 is a direct consequence of symmetry breaking of the typical 4-fold spin/valley degeneracy in graphene, which is a direct manifestation of the interaction-driven Fermi surface reconstruction which takes place when the CI gap is formed. While due to experimental limitations we could not measure our device in $B$-fields well above $B_0$, by continuity, all the LLs that above $B_0$ point away from the CNP, below $B_0$ point towards the CNP.

As we increase the magnetic field further from zero, the CI is continuously suppressed and vanishes at $B \sim 8$T, where the phase diagram becomes dominated by LLs. Strikingly, above $B > 24$T the $\nu = 2$ CI state reappears and grows continuously stronger up to $B = 30$ T (see Extended data C). Similar to $B = 0$ T, we also observe the emergence of a set of LLs nucleating from $\nu = 2$ and $\Phi_0$, with the sequence of $\nu_L = +2$, (+3), +4, (+5). Here the even fillings are much more strongly pronounced than the odd ones, as can be seen in Fig. 1e, which shows the quantum Hall traces at $B = 22$ T. In summary, the occurrence of an insulating state which is accompanied by a Fermi surface reconstruction at $B = 30$ T $= \Phi_0$ firmly establishes the existence of a re-entrant $\nu = 2$ CI in one magnetic flux quantum per moiré unit cell. Since the most dominant LL is here the $\nu_L = +2$, we also speculate that this LL can be interpreted as a correlated Chern insulator with a Chern number $C = 2$, in direct analogy to the $B = 0$ T case [26].

We further examine the full $\nu$-$B$ phase space of the flat bands. Figs. 2a and 2b show the respective Landau fan diagrams of $R_{xx}$ and $R_{xy}$ as a function of $\nu$ and $B$, where the observed LLs are schematically laid out in Fig. 2c. As discussed earlier, at $B = 0$ T we find clear signatures of CI states at $\nu = \pm 2$, +3, which are accompanied by Fermi surface reconstructions which are identified by new sets of LLs. Here originating from $\nu = \pm 2$ we observe a set of LLs with degeneracy 2. From the CNP we observe fully non-degenerate LLs, from $\nu = \pm 1$ we observe LLs with $\nu_L = \pm 3$, and from $\nu = \pm 3$ we observe LLs with $\nu_L = \pm 1$, in agreement with previous studies [26-28, 34].

Originating from $B_0$, we observe only one clearly pronounced CI at $\nu = 2$, and in contrast to the zero-field case we do not observe (in this sample) signatures of CIs at $\nu = -2$ and $\nu = 3$. We however observe a multitude of new sets of LLs which emerge from the different integer fillings $\nu$, showcasing that Fermi surface reconstructions are quite abundant at $\Phi_0$. Surprisingly it appears that all the Fermi surfaces have a fully lifted degeneracy, where from $\nu = +1$ we find LLs with the sequence $\nu_L = +1$, +2, +3, from $\nu = \pm 2$ we find LLs with $\nu_L = \pm 2$, ±3, ±4, ±5, and from $\nu = \pm 3$ we find LLs with $\nu_L = \pm 1$, ±2, ±3, ±4. This suggests that for both odd and even integers the spin and valley degeneracy have been lifted in full magnetic flux, which is very different than the zero-flux behavior. This can be attributed to the breaking of $C_{2z}T$ symmetry by magnetic flux [16, 29-31] – which lifts the degeneracy of the quasiparticles on top of the CIs, although quantitative predictions have not been made so far at odd fillings. Overall, we observe that the LLs converge at $\Phi_0$, indicating that $\Phi_0$ corresponds to one flux quantum per moiré unit cell. This implies that a well-defined moiré unit cell exists even in high magnetic fields. Furthermore, we also observe LLs which survive through the full range of magnetic field and connect two different integer fillings at zero flux and at $\Phi_0$ flux. Whereas [15] uses a strong coupling approach to analyze the ground states at $\Phi_0$, more theoretical work is required to study the interactions within the Hofstadter sub-bands at intermediate flux.

We briefly discuss the observed results. The interaction-driven gapped states at integer filling of the flat bands at zero magnetic field can give rise to several Chern bands due to the broken $C_{2z}T$ symmetry [15, 29, 32]. At $\Phi_0$, the $C_{2z}T$ symmetry is broken by the magnetic field, leading to different single-particle topology in the flat bands and different degeneracies in the many-body charge excitations atop the CI states, as discussed in Ref. [15]. The observed differences in the $B = 0$ T and $B = 30$ T LLs reflect the importance of symmetry and topology in determining the many-body phases. The observation of these LLs as well as the interaction-driven correlated insulators at certain integer fillings of the band at full magnetic flux $\Phi_0$ confirms the existence of the (theoretically confirmed) flat bands at $\Phi_0$ flux. The re-emergence of correlated insulating states is derived in Ref. [15], where it is shown that at full flux and with flat bands, the reappearance of full density Bloch-like bands (as opposed to fractional density Hofstadter bands that appear at rational flux) allows for the analytic construction of CI states.

Although interactions dominate the flat bands, the higher energy passive bands in MATBG have a larger bandwidth and are amenable to a single-particle Hofstadter analysis. A theoretical calculation of the BM model Hofstadter spectrum is possible at rational flux. Using the new gauge-invariant formalism of [33], we numerically compute the spectrum to very high accuracy between $\Phi = 0$ and $\Phi_0$. The results are shown in Fig. 3a where the largest gaps are marked by their Chern numbers, computed using Wilson loops and the Streda formula. We emphasize that our calculations are exact within the BM model, and do not rely on $\mathbf{k \cdot p}$ approximations near the Fermi surfaces. To access this regime experimentally, we have tuned the carrier density above $v > 4$ and populated the largely unstudied higher-energy dispersive bands of MATBG [26]. Fig. 3b shows the color plot of the longitudinal resistance $R_{xx}$ as a function of the normalized magnetic flux and carrier density. The strongest LLs, observable as lines with slope $1/v_L$, are schematically laid out in Fig. 3c with their filling factors $v$ and LL filling factors at $B = 0$ T, $(v, v_L)$, where we find a strong agreement (but also some disagreements) between the single-particle theory and the observed LLs. We now highlight the most important comparisons between theory and experimental data.

Ref. [12] demonstrated that at zero flux, the second through fifth bands of the BM model, counting from charge neutrality, form an elementary band representation and are forced to be connected by symmetry. Hence no gap is expected in the resistance data (Fig. 3b) between the fillings of $v = 4$ and $v = 20$ at 0 flux, although there is a Dirac point at filling $v \sim 12$ leading to a low density of states (see Extended data F). This can be seen in Fig. 3b from the deep blue conducting regions near $B = 0$, which are punctuated by the less conductive (lighter blue) region near $v = 12$. In magnetic flux, single-particle gaps may open at fractional fillings, as the full-density Bloch bands at zero flux are split into Hofstadter sub-bands. The Chern number $C$ of the gaps is given by the Streda formula: $N = pC \pmod{q}$, where $N$ is the number of Hofstadter subbands that have been filled, as measured from charge neutrality. At fractional filling ($N \neq 0 \pmod q$) the Chern number must be non-zero. When interactions, spin-orbit coupling, and the Zeeman effect are neglected, all bands are four-fold degenerate because of the spin and valley degeneracy.

From $\Phi = 0$ to $\Phi \sim 0.5\Phi_0$, there are three prominent features in the Wannier diagram (Fig. 3c) which appear at $v = 4$, 8, and 12 at $B = 0$ T. Near $v = 4$, we predict and observe positive slope LLs emerging from the band edge. These features have been well studied in Ref [26] and can be attributed to the low-energy Rashba point in the passive bands of the zero flux BM model. More interesting are the LLs which are connected to $v = 8$. Although there is no band edge at $v = 8$ in the BM model, the magnetic field breaks the $C_{2x}$ and $C_{2z}T$ symmetries that enforce the second and third bands to be connected. Thus, we can understand the LLs pointing to $v = 8$ as indicating a nascent band edge which is revealed by flux in precise agreement with Fig. 3a., which demonstrates strong gaps originating from $v = 8$, with Chern numbers 4, 8, 12, 16. We expect increased $R_{xx}$ in the regions where LLs of different slopes cross [26], as is observed in Fig. 3b near $v = 10$ where the $v = 8$, $v_L=4$ and $v = 12$, $v_L=-4$ LLs collide near $0.5\Phi_0$. We also mention that at $0.5\Phi_0$ flux, we observe very clean, high conductivity regions in Fig. 3b between the Chern gaps at $v = 8$, 10, 12, 14 and 16, corresponding to the metallic regions in the theoretically calculated Hofstadter spectrum (see Extended data F). marked by light blue bars in Fig. 3a and 3c. This demonstrates the high-quality Hofstadter bands within the sample at 15 T. The other prominent LLs emerge from $\Phi_0$ flux where the full flux quantization restores the zero-field Brillouin zone (Fig. 1b). Due to the breaking of $C_{2x}$ and $C_{2z}T$, all passive bands are gapped, leading to band edges at $v = 8$, 12 and 16 where new LLs converge as calculated in Fig. 3a. However, due to presumably the large contact resistance of the device in high fields, we could not identify these states.

Additionally, there are some clear LL features in Fig. 3b which are not explained by the single-particle Hofstadter calculations. The LLs in Fig. 3b and 3c with $v_L = 10$, 18, 22 (not divisible by 4) rely on interactions to break the spin-valley degeneracy. However, these LLs originate at $B = 0$ T from the charge excitations in the strongly interacting flat bands and appear to remain competitive many-body states even at large flux and high fillings. There are also $v = 4$ LLs with $v_L = 8$, 16 (divisible by 4) marked with dark grey lines in Fig. 3c which do not appear in our single-particle calculations, as well as a $\nu_L = 22$ many-body LL (also dark grey). Further work is necessary to characterize these states.

At last, we discuss the possibility of re-entrant superconductivity at $\Phi_0$ flux, which we have not observed in this study even though the device possesses a superconducting region on the hole doped side of the $\nu = -2$ CI at $B = 0$ T, with a high transition temperature of $T_c = 3$ K (see Fig. 1d and Extended data E for more details). While the exact nature of superconductivity in MATBG is still not clear, it is established that the ultra-high density of states in the flat bands is a key ingredient for its occurrence, and some theories tie it directly to the CI states. In direct analogy to the re-entrant behavior of the CI, it is possible that the flat bands in $\Phi_0$ flux could host re-entrant SC phases. However, the experimental parameters for its observation might be much more stringent than for the CIs. As the critical $B$-field of the SC is only about $B_c \sim 50$ mT, it is several orders of magnitude smaller than that for the CIs, and achieving such measurement accuracy at such high fields is challenging. Such small field can also be easily smeared out by twist-angle disorder, where at $\Phi_0$ flux a tiny twist-angle inhomogeneity of $\Delta\theta = 0.05°$ will give rise to a variation of the full flux value by $\Delta B = 60$ mT, which is enough to fully suppress a SC state. We hence propose the continued exploration of re-entrant SC in MATBG with ultra-homogeneous devices.

In summary, we have studied the behavior of MATBG in high magnetic field up to one flux quantum of the moiré unit cell $\Phi_0$, and have for the first time observed strong interaction-driven correlated insulating phases at integer fillings of the band. We also perform single particle Hofstadter calculations which predict a set of flat bands at full flux with different symmetry and band topology compared to 0 flux. These bands are unstable to the opening of correlated states by interactions [15].

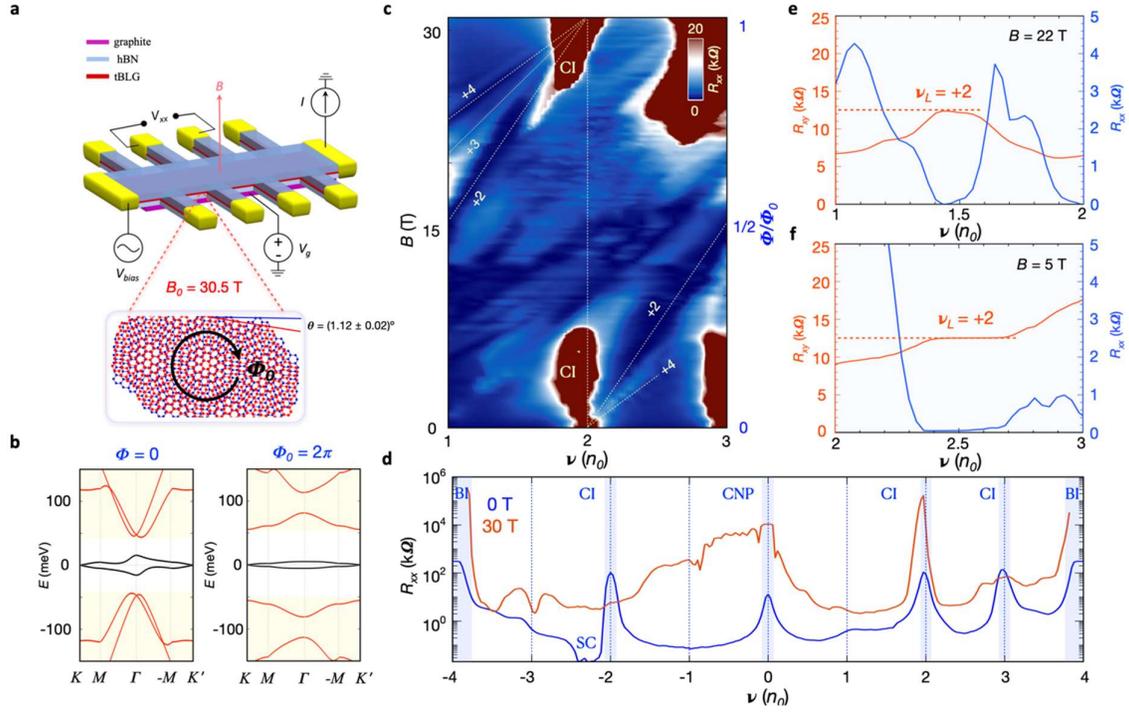

**FIG. 1. MATBG close to one flux quantum of the moiré unit cell $\Phi_0$. a)** Schematic of the MATBG device encapsulated with hBN, and the performed measurement schematic of the back gated four terminal $R_{xx}$ and $R_{xy}$ measurements. The twist angle of the device is $\theta = 1.12° \pm 0.02°$, which results in a magnetic field that corresponds to one flux quantum of the moiré unit cell $\Phi_0$ of $B_0 = 30.5$ T. **b)** Dispersive plot of the predicted low energy flat bands at zero and $\Phi_0$ magnetic flux. **c)** Color plot of $R_{xx}$ as a function of $\nu$ and $B$ measured at $T = 40$ mK around $\nu = 2$ filling of the band, showing the re-entrant CIs and Fermi surface reconstructions originating from $\Phi_0$. **d)** $R_{xx}$ vs. $\nu$ over the entire range of flat band at $B = 0$ T and $B = 30$ T, clearly showing the re-entrant CI at $\nu = +2$ close to the full magnetic flux quantum. **e)** and **f)** $R_{xx}$ and $R_{xy}$ as a function of $\nu$ at B = 22 T and $B = 5$ T, respectively, showing full quantization of the $\nu_L = +2$ LL gap.

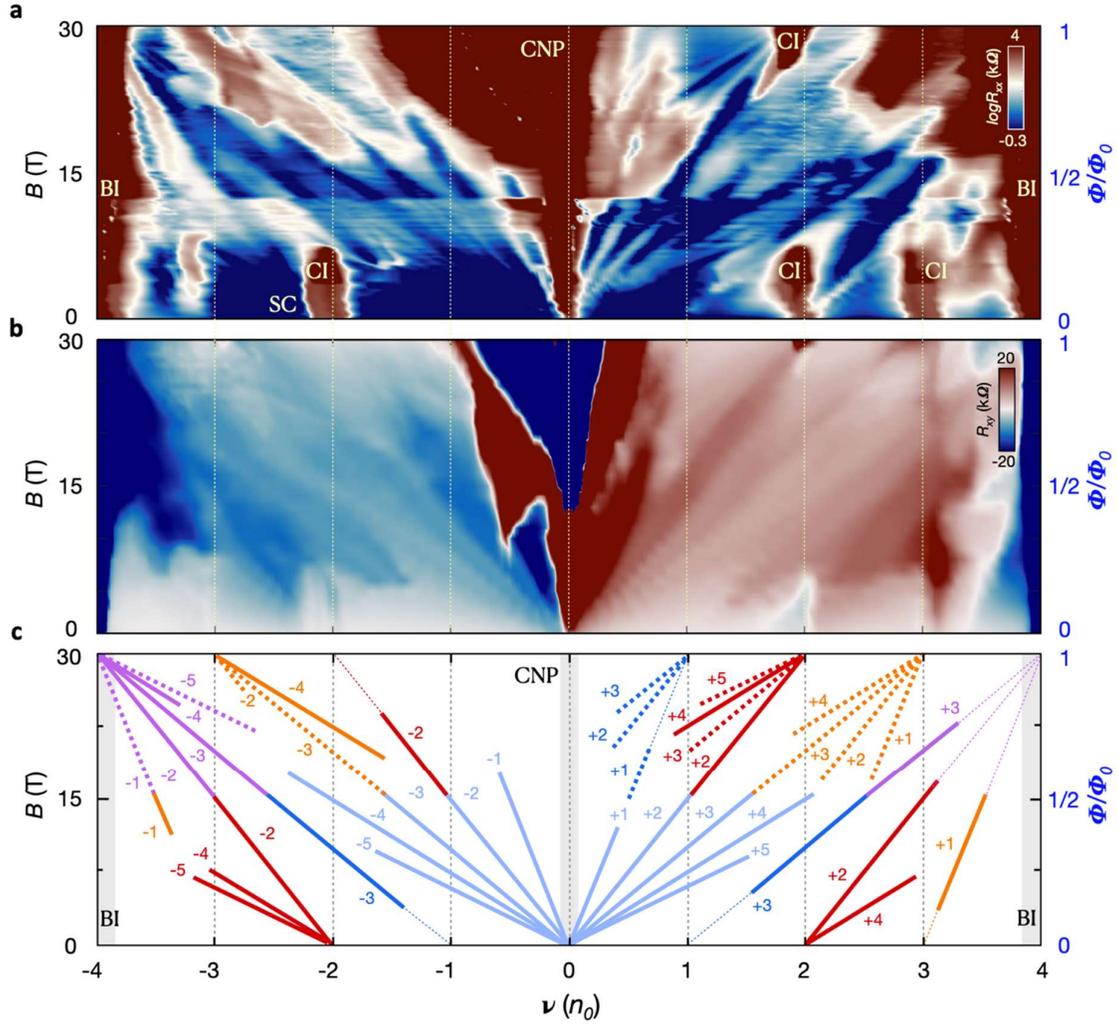

**FIG. 2. Landau fan diagrams from zero to one flux quantum of the moiré unit cell $\Phi_0$. a)** and **b)** show respectively the color plots of $R_{xx}$ and $R_{xy}$ as a function of $B$ and $\nu$, for the full magnetic phase space from $B = 0$ T to $B = 31$ T and $\nu$ from -4 to 4. **c)** Schematics of all the LL gaps emerging from different fillings of the band from both zero magnetic field and one flux quantum of the moiré unit cell $\Phi_0$. Light blue lines from CNP indicate the LL gaps with $\nu_L = \pm 1$, $\pm 2$, $\pm 3$, $\pm 4$, $\pm 5$. Dark blue lines indicate the LL gaps from $\nu = +1$ with $\nu_L = +3$ and $\nu_L = +1$, $+2$, $+3$ from zero-flux and $\Phi_0$ flux respectively and $\nu_L = -3$ from $\nu = -1$ at zero flux. Dark red lines correspond to LL gaps from $\nu = +2$ at both zero and $\Phi_0$ flux. Orange lines indicate the LL gaps that emerge from $\nu = \pm 3$ and purple lines correspond to LL gaps from $\nu = -4$. Solid lines mark well pronounced, quantized LL gaps, while dashed mark much weaker, non-quantized, gaps.

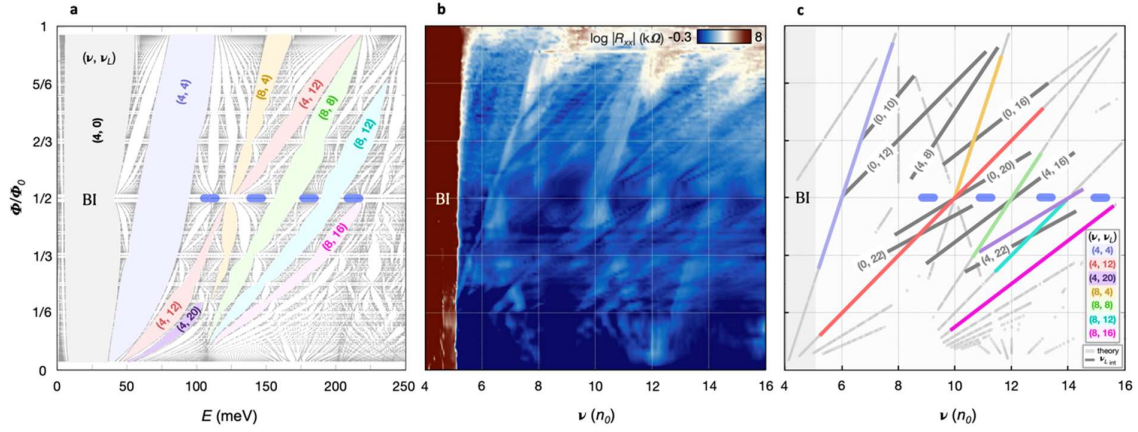

**FIG. 3**. **Hofstadter spectrum and LLs in the passive bands.** a) Calculated Hofstadter spectrum of the BM model for positive energy as a function of the magnetic flux $\Phi$ through the moiré unit cell. Different solid grey $(4, 0)$, blue $(4, 4)$, red $(4, 12)$, yellow $(8, 4)$, green $(8, 8)$, cyan $(8, 12)$, magenta $(8, 16)$ and purple $(4, 20)$ regions correspond to the evolution of dominant LL gaps. They are marked with the band filling and filling factor of the LLs $(v, v_L)$. b) Color plot of logarithmic $R_{xx}$ as a function of $B$ and $v$ for a very high range of carrier density up to $v = 16$. c) Schematics of (b) and the comparison from (a), where grey circles mark the theoretically predicted gaps from (a) and colored lines mark the strongest LL signatures in (b). The observed LLs from (b) are plotted with the same color code as is used for the corresponding LL gaps in (a), which shows clearly that many of the predicted gaps are well observed. In addition to the colored lines, dark grey lines correspond to LLs which are not predicted from the Hofstadter calculations. These levels do not come in multiples of 4 and are the result of strong interaction in the system ($v_{Lint}$). The horizontal blue bars denote metallic regions in (a) matching the high conductance regions (dark blue regions) observed in (b).

**Acknowledgements**

D.K.E. acknowledges support from the Ministry of Economy and Competitiveness of Spain through the 'Severo Ochoa 'programme for Centres of Excellence in R and D (SE5-0522), Fundacio Privada Cellex, Fundacio Privada Mir-Puig, the Generalitat de Catalunya through the CERCA programme, funding from the European Research Council (ERC) under the European Union's Horizon 2020 research and innovation programme (grant agreement no. 852927). B. A. B.'s work was primarily supported by the DOE Grant No. DE-SC0016239, the Schmidt Fund for Innovative Research, Simons Investigator Grant No. 404513, and the Packard Foundation. Z.-D. S. was supported by ONR No. N00014-20-1-2303, NSF-MRSEC No. DMR-1420541, Gordon and Betty Moore Foundation through Grant GBMF8685 towards the Princeton theory program. B.A.B. also acknowledges support from the European Research Council (ERC) under the European Union's Horizon 2020 research and innovation program (Grant Agreement No. 101020833). I.D. acknowledges support from the INphINIT 'laCaixa ' (ID 100010434) programme (LCF/BQ/DI19/11730030). K.W. and T.T. acknowledge support from the Elemental Strategy Initiative conducted by the MEXT, Japan (Grant Number JPMXP0112101001) and JSPS KAKENHI (Grant Numbers 19H05790, 20H00354 and 21H05233).

# Supplemental material: Observation of re-entrant correlated insulators and interaction driven Fermi surface reconstructions at one magnetic flux quantum per moiré unit cell in magic-angle twisted bilayer graphene

Ipsita Das[1], Cheng Shen[1], Alexandre Jaoui[1], Jonah Herzog-Arbeitman[2], Aaron Chew[2], Chang-Woo Cho[3], Kenji Watanabe[4], Takashi Taniguchi [5], Benjamin A. Piot[3], B. Andrei Bernevig[2] and Dmitri K. Efetov[1]*

1.  ICFO - Institut de Ciencies Fotoniques, The Barcelona Institute of Science and Technology, Castelldefels, Barcelona, 08860, Spain
2.  Department of Physics, Princeton University, Princeton, New Jersey 08544, USA
3.  Laboratoire National des Champs Magnétiques Intenses, Univ. Grenoble Alpes, UPS-INSA-EMFL-CNRS-LNCMI, 25 avenue des Martyrs, 38042 Grenoble, France
4.  Research Center for Functional Materials, National Institute for Materials Science, 1-1 Namiki, Tsukuba 305-0044, Japan
5.  International Center for Materials Nanoarchitectonics, National Institute for Materials Science, 1-1 Namiki, Tsukuba 305-0044, Japan

*E-mail : dmitri.efetov@icfo.eu

## A. Fabrication of MATBG device

Our MATBG samples were fabricated using a typical 'cut-and-stack 'method. The thickness of the hBN flakes used in these stacks was typically 10-15 nm. A single monolayer graphene flake was cut into two pieces using an AFM tip to prevent any accidental strain during the tearing process. A stamp of poly (bisphenol A carbonate) (PC)/polydimethylsiloxane (PDMS) mounted on a glass slide was used to pick up an hBN flake at 100–110°C. The hBN flake was then used to carefully pick up the first half of the pre-cut graphene piece from the Si++/SiO$_2$ (285 nm) substrate. The second layer of graphene was rotated by a target angle of $\theta$ = 1.1–1.2° and simultaneously picked up by the hBN/graphene stack from the last step at 100 °C. Subsequently, another hBN layer was picked up to completely cover the MATBG. Finally a few layers of graphite were picked up to complete the stack. In the end, the PC was melted at 180 °C and the full stack was dropped on an O$_2$-plasma cleaned Si++/SiO$_2$ chip. After cleaning the PC with chloroform, the final stack was closely examined using both optical and AFM images to make sure that the hBN and graphene edges were not aligned. The electrodes were made by CHF$_3$/O$_2$ plasma etching and deposition of Cr/Au (6nm/50nm) as the metallic edge contacts.

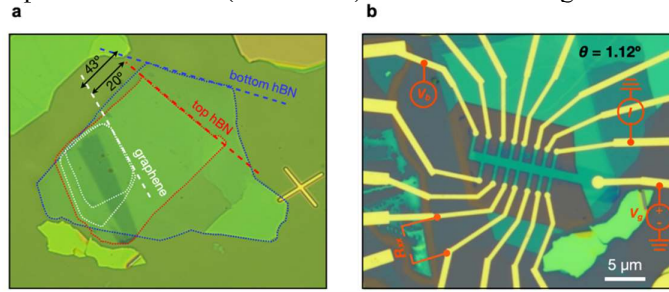

FIG. S1. Device fabrication and measurement scheme, a) Optical image of the twisted bilayer graphene encapsulated within two layers of hBN and graphite. Red, blue and white dotted lines show the edge of top-hBN, bottom-hBN and graphene respectively, and their misalignment by a large angle. b) Optical image of the fully fabricated device. The four probe measurement configuration has been shown in orange schematics. From transport measurements we calculated the twist angle of the device to be $\theta$ = 1.12°.

## B. Electrical transport measurements

All of the high magnetic field measurements were carried out in a dilution refrigerator with a base temperature of $T = 40$ mK. The maximum magnetic field available was $B = 31$ T. We have used a standard low-frequency lock-in technique (SR860 amplifiers) with an excitation frequency f = 13.111Hz to achieve a lower electron temperature in our measurements. Keithley 2400 source-meters were used to control the back-gate voltage. To characterize the superconducting state discussed later, a d.c. bias current was applied through a 1/100 divider and a 1MΩ resistor before combining with an a.c. excitation. The induced differential voltage was measured at the same frequency (13.111Hz) with the standard lock-in technique. We have used 100X amplifiers for all of our measurements in order to achieve a better signal-to-noise ratio since the high magnetic field induced substantial noise in the measurement.

The nano-fabrication process always incorporates some misalignment in the Hall bar geometry of the MATBG devices. At a very high magnetic field this misalignment of the probes becomes a dominating factor and incorporates a cross contribution to the longitudinal ($R_{xx}$) and Hall ($R_{xy}$) resistances. This prevents us from getting a perfectly isolated $R_{xx}$ and $R_{xy}$ measurement. The magnetic field symmetrization technique was used to extract these values with quantitative precision. Due to the magnetic field symmetrization property, the longitudinal resistances ($R_{xx}$) are symmetric under time reversal and are the same in positive or negative magnetic field, whereas the Hall resistances ($R_{xy}$) are antisymmetric and change sign for positive versus negative magnetic field. In order to get the exact value of these resistances, we measured the device in both the positive and negative polarities of the magnetic field (+$B$ and -$B$) and calculated $R_{xx}$ and $R_{xy}$ using these formulae:

$$R_{xx}(B) = (R_{meas}(+B) + R_{meas}(-B))/2$$

$$R_{xy}(B) = (R_{meas}(+B) - R_{meas}(-B))/2$$

Due to experimental difficulties, we could not measure the device in the negative values of magnetic field for the full $B$ phase space up to -31 T. However, we were able to perform measurements for certain values of magnetic field. In Fig. 1e and 1f, the Hall resistance $R_{xy}$ was antisymmetrized in order to get perfect quantization at $\nu_L = +2$.

## C. Magnetic field dependence of the correlated insulators

Fig. S2a and S2b show the magnetic field dependence of the correlated insulator at $\nu = +2$ for both zero flux and $2\pi$ flux. A perpendicular magnetic field has a direct effect on this insulating state. A small magnetic field of $B = 5$ T suppresses the resistance of this state and weakens the insulator as shown in Fig. S2c. The insulator is destroyed upon further increasing the magnetic field ($B = 8$ T) where the conductivity rapidly increases.

We have observed qualitatively similar behavior for the correlated insulator near $2\pi$ flux at $\nu = +2$. Fig. S4b shows a color plot of $R_{xx}$ close to $\nu = +2$ and $\Phi = 2\pi$, from $B = 22$ T to $B = 30$ T. The highly resistive state visibly fades upon lowering the magnetic field away from the flux quantum $\Phi_0 = 31$ T. $R_{xx}$ has a maximum value at $B = 31$ T, decreases with magnetic field, and finally obtains high conductance at $B = 26$ T as shown in Fig. S2d. This same qualitative trend of the correlated insulator at $\nu = +2$ for both zero flux and $\Phi_0$ suggests the two correlated insulators possess a similar origin. Due to experimental limitations, we could not apply an external magnetic field higher than $\Phi_0$ or $B = 31$ T.

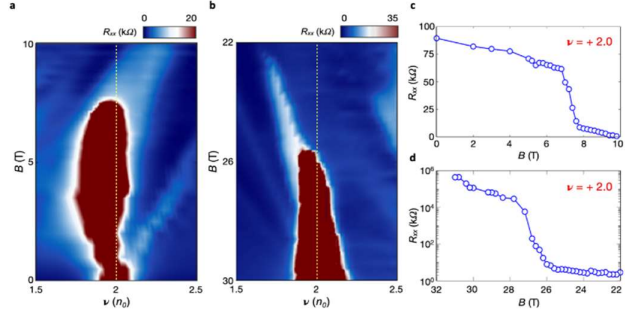

FIG. S2. Magnetic field dependence of the correlated insulator at $\nu = +2$. a) Color plot of $R_{xx}$ as a function of $B$ and $\nu$ close to zero flux. Deep red regions indicate the presence of a strong insulator. b) Another color plot of $R_{xx}$ as a function of $B$ and $\nu$, again depicting the correlated insulator, but for $B$ close to $\Phi_0$ flux. c) Evolution of the $\nu = +2$ state with magnetic field ranging from $B = 0$ T to $B = 10$ T. d) Evolution of the $\nu = +2$ state, but for high magnetic field from $B = 31$ T to $B = 22$ T (note the reversed axis). $R_{xx}$ is plotted in logarithmic scale.

## D. Temperature dependence of the correlated insulators

At zero magnetic field, we have clearly observed correlated insulators at $\nu = \pm2$, $+3$. We have characterized the temperature dependence of these states and calculated the activation gap via Arrhenius fitting. The blue, green and yellow line plots in Fig. S3a show the conductance ($G_{xx}$) of the $\nu = +2$, $+3$ and $-2$ states, respectively, as a function of $1/T$ at zero magnetic field. From the Arrhenius fitting, we calculated their gaps to be $\Delta = 0.22$ meV ($\nu = +2$), $0.09$ meV ($\nu = +3$) and $0.21$ meV ($\nu = -2$) respectively. We have also observed correlated insulating states at $B = 31$ T. Fig. S3b shows the conductance as a function of $1/T$ at 24 T for states from $\nu = +2$ to $\nu = +2.1$. Fig. S3c also indicates the insulating trend for the $\nu = +3$ state. Due to experimental limitations at high magnetic field, we could not measure the full temperature dependence of the highly resistive insulating states at $\nu = +2$, $+3$ close to full flux. However, an insulating trend was clearly observed for these states in the available data up to 1.2 K. An exact extraction of the temperature activated gaps of these states requires future measurements at high magnetic field in a variable temperature insert (VTI).

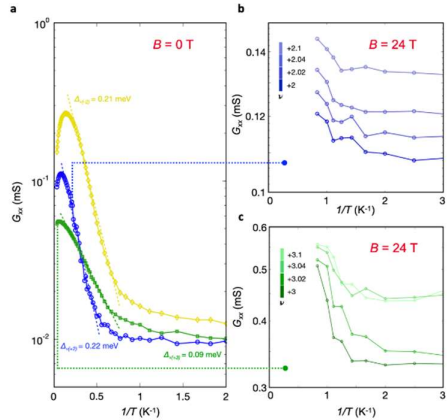

FIG. S3. Temperature dependence of the CIs. a) Conductance ($G_{xx}$) as a function of $1/T$ for the correlated insulating states at $\nu = +2$ (blue), $+3$ (green), and $-2$ (yellow) at zero magnetic field. From the Arrhenius fit, the extracted temperature-activated gaps are $\Delta = 0.22$ meV, $0.09$ meV and $0.21$ meV, respectively. Blue and green dotted lines guide the eye to the corresponding states at $\Phi_0$ flux. b) $G_{xx}$ as a function of $1/T$ for states at $\nu = +2$, $+2.02$, $+2.04$ and $+2.1$ at $B = 24$ T. Lighter shades of blue indicate higher values of $\nu$. c) $G_{xx}$ as a function of $1/T$ for states at $\nu = +3$, $+3.02$, $+3.04$ and $+3.1$ at $B = 24$ T. Lighter shades of green indicates higher value of $\nu$.

## E. Superconductivity in zero field

We have observed a strong superconducting dome close to half-filling of the flat bands at zero magnetic field in the hole side. Fig. S4a shows the longitudinal resistance $R_{xx}$ as a function of temperature $T$ and filling $v$. The dark red region at $v = -2$ confirms the well-known correlated insulating state at half filling. When the band is slightly doped away from this insulating state, $R_{xx}$ vanishes. The dark blue region ($v$ = -2.1 to -2.32) in Fig. S4a corresponds to the SC dome. The typical temperature dependence of the resistance ($R_{xx}$) at the optimal doping of the SC dome ($v = -2.26$) shows a clear transition to a zero resistance state. In Fig. S2b we calculated the transition temperature to be $T_c = 3$ K. Fig. S4c shows the typical d.c. IV measurement of the superconductor. It has a very high critical current, $I_c = 400$ nA, at the base temperature of our fridge ($T = 40$ mK) indicating the high quality of the device.

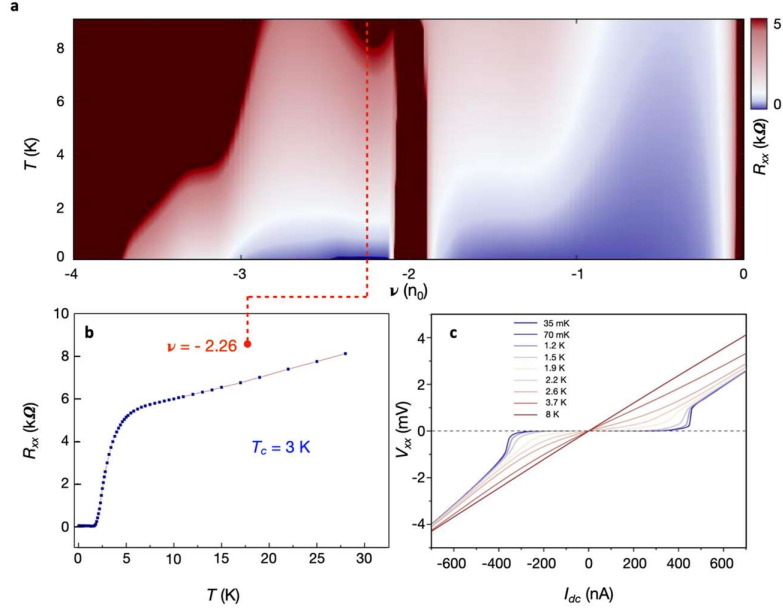

FIG. S4. Superconductivity. a) Temperature dependence of $R_{xx}$ as a function of $v$ for the hole side of the flat band. The dark red regions at $v = 0$, -2, -4 correspond to the charge neutrality point, a correlated insulator, and a band insulator, respectively. The dark blue region left of $v$ = -2 shows the superconducting dome. b) $R_{xx}$ as a function of $T$ for the optimal doping of the SC at $v = -2.26$, yielding a critical temperature $T_c = 3$ K. c) DC current versus voltage plots for the SC at different temperatures from $T = 35$ mK to 8 K.

## F. Theoretical calculation of single-particle spectrum in magnetic flux

In this Appendix we discuss the single-particle spectrum of twisted bilayer graphene in a magnetic field computed using the Bistrizter-MacDonald model [1]. For completeness, we include the zero-magnetic field spectrum in Fig. S5a, where we highlight the two flat bands near zero energy (black) and the four connected lowest passive bands above and below charge neutrality (blue). Fig. S5b and c show that density of states (DOS) and filling $n(E)$ respectively. In the Main Text, we compare some features of the low magnetic field resistance data (Main Text Fig. 3b) to the zero flux band structure which can be understood from a Landau level approximation of the $\mathbf{k} \cdot \mathbf{p}$ spectrum. However in larger fields, a full Hofstadter calculation is necessary to determine the spectrum and topology.

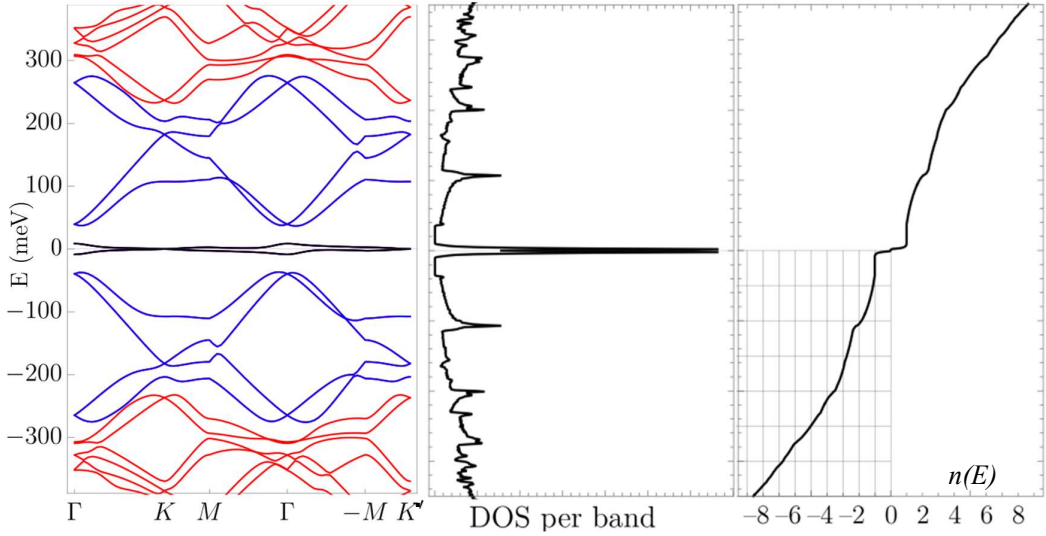

FIG S5. BM model with common energy ($E$) axis. **a)** We show the bands the single-particle bands of the BM model $\theta = 1.12^o$ in zero flux. The nearly flat bands at charge neutrality are black, the lowest four connected passive bands are blue, and the sixth and higher bands are red (not considered in this work). **b)** We show the normalized density of states (arbitrary units). **c)** We compute the integrated DOS or filling $n(E)$ as a function of the energy. The 50 meV hatching in the bottom left corner is for visual ease of comparison with energies are integer filling of the bands. The filling is reported *without* the spin-valley degeneracy, so $n = 1$ corresponds to $\upsilon = 4$.

The band structure of the BM model in $2\pi$ magnetic flux per unit cell was calculated in Ref. [2], wherein the authors developed a gauge-invariant formalism to study the single-particle spectrum, topology, and interacting states. By employing a judicious choice of basis states that are eigenstates of both magnetic translation operators (which commute at $2\pi$ flux), they were able to block-diagonalize the continuum Hamiltonian by momentum, allowing for calculations of band structures and Wilson loops. The momentum basis states at flux $\Phi = 2\pi$ read

$$\psi_{\mathbf{k},m}(\mathbf{r}) = \frac{1}{\sqrt{\mathcal{N}(\mathbf{k})}} \sum_{R_1,R_2 \in \mathbb{Z}} e^{-i\mathbf{k}\cdot\mathbf{R}} T_1^{R_1} T_2^{R_2} w_m(\mathbf{r}), \quad T_1 T_2 = e^{i\Phi} T_2 T_1, [T_1, H] = [T_2, H] = 0.$$

Here $\mathbf{k}$ is momentum and $\mathbf{r}$ is position, $w_m(\mathbf{r})$ is the wavefunction for the $m$th Landau level centered at the origin, and $T_1$, $T_2$ are the magnetic translation operators. Notably, the magnetic translation operators $T_i$ commute at $\Phi = 2\pi$, so, as is the case at zero field, the spectrum is diagonalized into Bloch-like bands with density one electron per moiré unit cell. The $m$th momentum basis states are in essence linear combinations of the $m$th Landau level translated by all lattice vectors $\mathbf{R}$, and appropriately normalized by $\mathcal{N}(\mathbf{k})$, for which an expression may be found in Ref[2]. When expressed in this basis, the Bistritzer-MacDonald model in $2\pi$ flux becomes a relatively simple Hamiltonian involving ratios of Jacobi theta functions. The Landau level index $m$ is chosen to range from to $n_{\text{landau}}$, the Landau level cutoff. Increasing $n_{\text{landau}}$ improves the accuracy of the eigenvalues and eigenvectors.

To treat twisted bilayer graphene at rational flux $\Phi = 2\pi p/q$, where $p$ and $q$ are coprime integers, the authors have developed an extension of this formalism. One may map the rational flux Hamiltonian to a $2\pi$ flux Hamiltonian by a redefinition of the lattice vectors, shrinking one lattice vector by a factor of $p$ (at the cost of the Hamiltonian no longer being translation symmetric with respect to this smaller lattice vector) and extending the other by a factor of $q$. This modified unit cell encloses $2\pi$ flux, and so the results of Ref. [2] may be applied with only minor modifications. It is important to note that the resulting Hofstadter sub-bands have lower density: they have one electron per $q$ moiré unit cells. We will explain in detail the irreducible representations of the rational flux magnetic translation operators, the calculation of topology via Wilson loops, and the magnetic space group symmetry analysis in a forthcoming paper Ref. [3].

Because the Hamiltonians are block-diagonalized by momentum, they are relatively small in dimension, although the Hamiltonians are dense. It is not numerically taxing to build a large Hofstadter spectrum reaching as high as $q = 40$, with $p$ ranging from 1 to $q$. Each Hamiltonian for a given momentum has dimension $\approx 4pn_{landau}$ x $4pn_{landau}$, where $n_{landau}$ is the Landau level cutoff. Even for small $n_{landau}$, for example 50, we find the gauge-invariant rational flux technique works very well and is quite accurate for most values of $\Phi$; it is only for high $q$ that the Hamiltonian loses accuracy, requiring an increase in $n_{landau}$ to compensate. The Hamiltonian is diagonalized over a mesh of points in the magnetic Brillouin zone, yielding a Hofstadter spectrum shown in Fig. S6. We neglect the Zeeman splitting, which is less than 2meV. Traces of the inaccuracy at high $q$ can be seen in the hairs that split off from the bands in Fig. S6b, at $p/q \approx .12$; for higher $q$ these inaccuracies are quickly remedied by choosing higher values of the Landau level cutoff.

Gaps in the Hofstadter spectrum correspond to single particle insulating states. Unlike 0 and $2\pi$ flux, at rational flux single-particle gaps occur also at fractional fillings. To see this, note that at rational flux $\Phi = 2\pi p/q$, the magnetic translation operators $T_1$, $T_2$ do not commute with one another. However, the operators

$$[T_1, T_2^q] = 0, [T_1, H] = 0, [T_2^q, H] = 0$$

do commute, implying the Hamiltonian can be diagonalized in a magnetic unit cell enlarged by a factor of $q$. Thus the magnetic Brillouin zone is reduced by a factor of $q$ from that at zero field. At 0 and $2\pi$ flux, a filling of $\nu = 4$ (4 for spin and valley degeneracy) implies that one band is completely filled. At rational flux, a filling of $\nu = 4$ fills $q$ Hofstadter bands (per valley per spin, a total of $4q$ bands if opposite spin and valley count as separate bands); this is because the Brillouin zone is smaller, so a higher numbers of bands are filled to have the same electron density. We searched the Hofstadter spectrum for gaps greater than our chosen cutoff of 6 meV and selected those that occurred at precisely an integer number of Hofstadter bands (per valley per spin) away from charge neutrality.

The Chern number of these gaps are computed via Wilson loops and the Streda formula in Fig. 3a of the Main Text, and we plotted the gaps in the B vs. $\nu$ Wannier diagram in Fig. 3c of the Main Text. Each gap is plotted as a grey circle, and these circles arrange into lines whose slope corresponds to the Chern number of the band gap. For the passive bands, the single-particle predictions of Chern insulators well agree with the experimental data (see Fig. 3b of the Main Text). In the flat bands however, the observed Landau level spectrum (Fig. 1 of the Main Text) is completely at odds with the single-particle predictions of Fig. S6b. This is evidence that the single-particle picture is inadequate and a strong coupling approach is necessary to describe the flat bands.

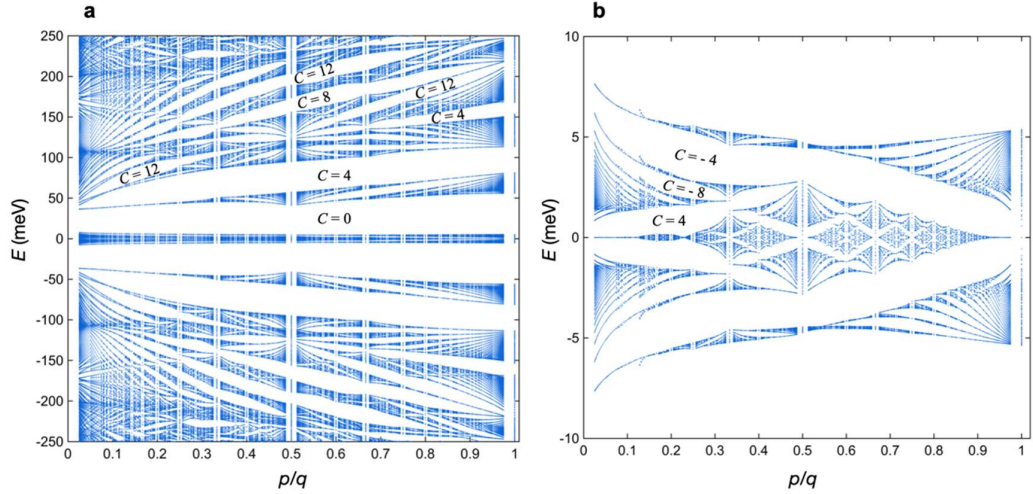

FIG S6. Hofstadter Spectrum. **a)** We show the bands, the Hofstadter gaps, and their Chern numbers (including spin-valley degeneracy) over $\pm 250$meV at $\theta = 1.12^o$ for flux $2\pi p/q$ where $2\pi$ flux = 31T. The flat bands remain gapped for all flux, while the passive bands develop a complex structure with many gaps larger than 6meV. See Ref. [4] for a discussion of the topology of the flat bands. **b)** We focus on the flat bands near zero energy. In this regime, the ~20meV Coulomb interaction dominates over the kinetic energy, suppressing any single-particle signatures. The experimentally observed Landau level fans thus must arise from the strong coupling ground states as in Ref. [4]. Our primary observation is that the single-particle bands undergo significant restructuring from 0 to $2\pi$ flux. At $2\pi$ flux, the original Brillouin zone is restored and the bandwidth of the flat bands is very similar to the zero flux bands.